\documentclass[runningheads]{svmult}

\usepackage{makeidx}   
\usepackage{graphicx}  
\usepackage{subeqnar}  
\usepackage{multicol}  
\usepackage{physprbb}  
\makeindex             

\usepackage{amsfonts}   
\begin{document}
\title*{The Importance of Boundary Conditions\protect\newline 
in Quantum Mechanics}
\toctitle{The Importance of Boundary Conditions\protect\newline 
in Quantum Mechanics}
%
%
\titlerunning{The Importance of Boundary Conditions in Quantum Mechanics}
%

\author{Rafael de la Madrid}
\authorrunning{Rafael de la Madrid}

\institute{Institute for Scientific Interchange (ISI), 
Villa Gualino \\ 
Viale Settimio Severo 65, I-10133, Torino, Italy \\
E-mail: \texttt{rafa@isiosf.isi.it}}


\maketitle              

\begin{abstract}
\noindent We discuss the role of boundary conditions in determining the 
physical content of the solutions of the Schr\"odinger equation. We study the 
standing-wave, the ``in,'' the ``out,'' and the purely outgoing boundary 
conditions. As well, we rephrase Feynman's $+\I \varepsilon$ prescription as a 
time-asymmetric, causal boundary condition, and discuss the connection of 
Feynman's $+\I \varepsilon$ prescription with the arrow of time of Quantum 
Electrodynamics. A parallel of this arrow of time with that of Classical 
Electrodynamics is made. We conclude that in general, the time evolution of 
a closed quantum system has indeed an arrow of time built into the propagators.
\end{abstract}

\section{Introduction}

In physics, dynamical equations often have a differential form and are solved
under various boundary conditions. This is also the case in Quantum Mechanics,
whose dynamics is encoded by the Schr\"odinger equation. The purpose
of this contribution is to discuss, in a somewhat sketchy way, how boundary 
conditions imposed upon the Schr\"odinger equation determine the physical
content of its solutions. And vice versa: in order to obtain the solutions of 
the Schr\"odinger equation that describe a given physical situation, boundary 
conditions that fit the physical situation must be imposed upon the 
Schr\"odinger equation. 

In the non-relativistic domain, the time-independent Schr\"odinger equation 
is realized, in the position 
representation, as a second-order differential equation. If the potential
is simple enough, we can exactly solve the differential equation by means of 
the Sturm-Liouville theory~\cite{DIS,BRANDAS1,BRANDAS}. This theory yields an 
eigenfunction expansion, a 
unitary operator that diagonalizes the Hamiltonian, and a direct integral 
decomposition. In its turn, the direct integral decomposition yields, along 
with some physical requirements, a Rigged Hilbert Space (RHS). The 
eigenfunction expansion, the unitary operator that diagonalizes the 
Hamiltonian, the direct integral decomposition, and the RHS contain much of 
the spectral and physical informations of the Hamiltonian. For convenience, we 
shall refer to them as the RHS properties (abbreviated RSP) associated to the 
Hamiltonian~\cite{RSP}. At first sight, it may seem that the 
Schr\"odinger equation 
corresponding to a given Hamiltonian generates just one single RSP. However, 
this is not necessarily so. There are certain types of boundary conditions 
that, when imposed upon the Schr\"odinger equation, yield different RSPs. To 
be more precise, the standing-wave boundary condition (to be defined below), 
and the ``in'' and ``out'' boundary conditions of the Lippmann-Schwinger 
equation generate three different RSPs.

The Gamow vectors are the state vectors of 
resonances~\cite{GAMOW,SIEGERT,MONDRAGON,FERREIRA,AJP,BG,DIS}. Like
the standing-wave and the Lippmann-Schwinger eigensolutions,
they solve the Schr\"odinger equation. At infinity, however, the Gamow 
eigenfunctions satisfy the purely outgoing boundary condition. This condition 
selects the (complex) resonance spectrum of the Schr\"odinger equation.

The time-dependent Schr\"odinger equation is time 
symmetric, the reason for which it is generally believed that Quantum 
Mechanics is time symmetric~\cite{HALLIWEL}. And yet 
the time evolution of individual atoms or subatomic 
particles seems to have some directness. For example, imagine that we want to
compute the probability for a particle to go from an initial 
space-time location $({\bf x},t)$ to a final space-time location 
$({\bf x}',t')$. Our basic notions of causality dictate that this probability 
be zero when $t'<t$. However, as far as the Schr\"odinger equation is 
concerned, this probability is also non-zero when $t'<t$. In order to 
obtain the causal result, we have to use the retarded propagator
$G^+({\bf x},t;{\bf x}',t')$ (see for example~\cite{SCHIFF}). This
retarded propagator automatically yields a causal probability, because it
vanishes when $t'<t$:
\begin{equation}
       G^+({\bf x},t;{\bf x}',t')=0 \, ,  \quad t'<t \, .
       \label{propag}
\end{equation}
Therefore, the time evolution given by $G^+({\bf x},t;{\bf x}',t')$ has an
arrow of time: a particle travels forward in time, never backward. As is well 
known, the time evolution given by the (e.g., retarded) propagator is
equivalent to the time-dependent Schr\"odinger equation subject to an 
(e.g., retarded) causal boundary condition. Hence, the arrow of time
built into $G^+({\bf x},t;{\bf x}',t')$ stems ultimately from causal boundary 
conditions.

Quantum Electrodynamics (QED) provides a glaring example of how the 
propagators carry an arrow of time. In QED, the Feynman propagator is 
constructed by imposing Feynman's $+\I \varepsilon$ prescription upon the time 
evolution of particles and antiparticles~\cite{FEYNMAN}. Particles travel 
forward in time, whereas antiparticles ``travel backward'' in time. Clearly, 
this prescription builds an arrow of time into the Feynman propagator. But 
note that this arrow of time is introduced by means of Feynman's 
$+\I \varepsilon$
prescription, which is a boundary condition. Thus, the arrow of time of the 
Feynman propagator also stems ultimately from causal boundary conditions.

\vskip0.4cm

The organization of this contribution is as follows: 

\vskip0.4cm

({\it i}) In order to obtain the standing-wave eigensolution 
$\langle r|E\rangle$,
the ``in'' Lippmann-Schwinger eigensolution $\langle r|E ^+ \rangle$, and the 
``out'' Lippmann-Schwinger eigensolution $\langle r|E ^- \rangle$, we shall 
impose upon the Schr\"odinger equation the standing-wave, the ``in,'' and 
the ``out'' boundary condition, respectively. Each of the eigensolutions
$\langle r|E\rangle$, $\langle r|E ^+ \rangle$, $\langle r|E ^- \rangle$
yields an RSP of its own. Hence, we shall conclude that each of the 
eigensolutions
$\langle r|E\rangle$, $\langle r|E ^+ \rangle$, $\langle r|E ^- \rangle$
has a physical content of its own.

\vskip0.4cm

({\it ii}) It will be apparent that what makes
$\langle r|E\rangle$, $\langle r|E ^+ \rangle$, $\langle r|E ^- \rangle$
different from each other is the boundary conditions that they satisfy. More 
precisely, what makes them different from each other is their asymptotic
behavior at infinity.  

\vskip0.4cm

({\it iii}) We will relate the asymptotic behavior at infinity of
$\langle r|E\rangle$, $\langle r|E ^+ \rangle$, $\langle r|E ^- \rangle$
with their analytical dependence on the energy, and see why applying the
Sturm-Liouville theory to these eigenfunctions yields three different RSPs.

\vskip0.4cm

({\it iv}) When two eigenfunctions have just a different normalization, they 
generate the same RSP. A criterion to check whether or not two 
eigensolutions lead to different RSPs (i.e., whether or not two eigensolutions
differ from a normalization factor) is provided. We shall apply the 
criterion to $\langle r|E\rangle$, $\langle r|E ^+ \rangle$, 
$\langle r|E ^- \rangle$ and see (as expected) that they are not a 
normalization of each other.

\vskip0.4cm

({\it v}) We shall compare the boundary conditions satisfied by the Gamow 
vectors with those satisfied by 
$\langle r|E\rangle$, $\langle r|E ^+ \rangle$, and
$\langle r|E ^- \rangle$. More precisely, we shall compare the 
purely outgoing boundary condition with the standing-wave,
``in,'' and ``out'' boundary conditions, and see why the purely
outgoing boundary condition determines the physical content of the Gamow
vectors.

\vskip0.4cm

({\it vi}) We shall discuss the time asymmetry of QED. We shall refer to this 
time asymmetry as the QED arrow of time. We shall see that the QED arrow of 
time is built into Feynman's propagator. A parallel of the QED arrow of time 
with that of Classical Electrodynamics will be made. We shall conclude that, 
in general, there exists an arrow of time at the microscopic level, and that 
this arrow of time arises from the imposition of a time-asymmetric, causal 
boundary
condition upon the (time-symmetric) Schr\"odinger equation. Because solving
the Schr\"odinger equation subject to a time-asymmetric boundary condition 
necessarily involves the construction of a propagator, we shall conclude that 
the quantum-mechanical arrow of time is built into the propagators.

\section{Boundary Conditions upon the Time-Independent Schr\"odinger Equation}
\label{sec:tise}

We proceed now to see how boundary conditions affect the behavior of the 
solutions of the time-independent Schr\"odinger equation. Rather than working
in a general fashion, we shall use the spherical shell potential as an 
illustrative example. Generalizations to more complicated potentials are 
straightforward.

Consider the spherical shell potential of height $V_0$,
\begin{equation}
	V(\vec{x})=V(r)=\left\{ \begin{array}{ll}
                                0   &0<r<a  \\
                                V_0 &a<r<b  \\
                                0   &b<r<\infty \, .
                  \end{array} 
                 \right. 
	\label{potential}
\end{equation}
If we restrict ourselves to the zero angular momentum case, then
the spherical shell Hamiltonian acts, in the radial position representation, as
the following formal differential operator:
\begin{equation}
       h \equiv -\frac{\hbar ^2}{2m} \frac{\D ^2}{\D r^2}+V(r) \, .
      \label{doh}
\end{equation}
The time-independent Schr\"odinger equation (for zero angular momentum) 
reads
\begin{equation}
      \left( -\frac{\hbar ^2}{2m} \frac{\D ^2}{\D r^2}+V(r) \right) \sigma (r;E)=
      E\sigma (r;E) \, .
      \label{rSe0}
\end{equation}

Our objective in this section is to solve (\ref{rSe0}) subject to
various boundary conditions, and to analyze the physical content of
both the solutions and the boundary conditions. We shall study three 
cases: the standing-wave, the Lippmann-Schwinger, and the Gamow eigensolutions.

\subsection{Standing-Wave Eigenfunctions}
\label{sec:dk}

We first study the standing-wave eigenfunctions. To obtain them, we solve
(\ref{rSe0}) under the following boundary conditions:
\begin{subeqnarray}
       &&\sigma (0;E)=0 \, , \label{raogr}  \\
       &&\sigma (r;E) {\rm \ is \ continuous \ at \ } r=a {\rm \
      and \ at \ } r=b \, , \label{bocchiata1} \\
      &&\frac{\D }{\D r}\sigma (r;E) {\rm \ is \ continuous \ at \ } 
       r=a {\rm \ and \ at \ } r=b \, . \label{bocchiata4}
\end{subeqnarray} 
The eigensolution of (\ref{rSe0}) that satisfies 
(\ref{raogr})--(\ref{bocchiata4}) is given by the regular solution:
\begin{equation}
      \chi (r;E)\equiv \chi (r;k)=\left\{ \begin{array}{lll}
               \sin (kr) \quad &0<r<a  \\
               {\cal J}_1(k)\E ^{\I Qr}
                +{\cal J}_2(k)\E ^{-\I Qr}
                 \quad  &a<r<b \\
               {\cal J}_3(k) \E ^{\I kr}
                +{\cal J}_4(k)\E ^{-\I kr}
                 \quad  &b<r<\infty \, ,
               \end{array} 
                 \right. 
             \label{chi}
\end{equation}
where
\begin{equation}
      k=\sqrt{\frac{2m}{\hbar ^2}E \,} \, , \quad 
      Q=\sqrt{\frac{2m}{\hbar ^2}(E-V_0) \,} \, .
\end{equation}
The coefficients ${\cal J}_1(k)$--${\cal J}_4(k)$ in (\ref{chi}) are such 
that $\chi (r;E)$ satisfies (\ref{bocchiata1}) and (\ref{bocchiata4}). The 
expressions of ${\cal J}_1(k)$--${\cal J}_4(k)$ can be easily calculated 
(they can also be found, for example, in~\cite{DIS}).  

The regular solution $\chi (r;E)$ is not $\delta$-normalized. In order to 
$\delta$-normalize it, we need the spectral measure
\begin{equation}
      \varrho (E)\equiv \varrho (k)=
      \frac{1}{4\pi}\, 
      \frac{2m/\hbar ^2}{k}
      \, \frac{1}{ |{\cal J}_4(k)|^2}   \, .
\end{equation} 
Multiplying $\chi (r;E)$ by the square root of this spectral measure yields
the $\delta$-normalized eigensolution of (\ref{rSe0}) that satisfies
the boundary conditions (\ref{raogr})--(\ref{bocchiata4}): 
\begin{equation}
      \langle r|E\rangle \equiv \sqrt{\varrho (k)} \, \chi (r;k) \, .
      \label{dnes}
\end{equation}
The $\delta$-normalization of $\langle r|E\rangle$ is to be understood in the 
following sense:
\begin{equation}
       \int_0^{\infty}\D r \, \langle E|r\rangle \langle r|E'\rangle =
         \delta (E-E') \, , \quad E, E'\in [0, \infty ) \, ,
       \label{detanor}
\end{equation}
where $\langle E|r\rangle$ is the complex conjugate of 
$\langle r|E\rangle$. Note that although $\langle r|E\rangle$ is also defined 
for complex energies, we have restricted $E$ to $[0, \infty )$, because the 
(Hilbert space) spectrum of the Hamiltonian (\ref{doh}) is the positive 
real line. 

At infinity, the eigensolution (\ref{dnes}) is a linear combination of
\begin{equation}
   \sqrt{\varrho (k)}\, {\cal J}_4(k) \, \E ^{-\I kr} \, ,
   \label{inc}
\end{equation}
which is an incoming spherical wave of amplitude 
$\sqrt{\varrho (k)}\, {\cal J}_4(k)$, and 
\begin{equation}
  \sqrt{\varrho (k)}\, {\cal J}_3(k)\, \E ^{\I kr} \, , 
   \label{outc}
\end{equation}
which is an outgoing spherical wave of amplitude 
$\sqrt{\varrho (k)}\, {\cal J}_3(k)$. It is easy to see that when $E$ is real,
(\ref{inc}) is the complex conjugate of (\ref{outc}). Thus, far away from the 
potential region, $\langle r|E\rangle$ is the linear combination of an 
incoming spherical wave and its complex conjugate. This behavior is very much 
like that of a sinusoidal function -- hence the name standing-wave solution 
for the eigenfunction $\langle r|E\rangle$.

As shown in~\cite{DIS}, the eigenfunctions (\ref{dnes}) generate, by 
means of the Sturm-Liouville theory,\footnote{Our basic reference on the  
Sturm-Liouville theory is~\cite{DUNFORD}. Illustrative applications of
the Sturm-Liouville theory can be found, for example, 
in~\cite{DIS,BRANDAS1,BRANDAS,JPA,CSF,FP}.} an RSP; that is, the 
$\langle r|E\rangle$ generate an eigenfunction 
expansion, a unitary operator $U$ that diagonalizes the Hamiltonian, a 
direct integral decomposition, and an RHS
\begin{equation}
       \mathbf \Phi \subset {\cal H} \subset {\mathbf \Phi}^{\times} \, .
      \label{rhssw}
\end{equation}
(The explicit form of the RSP generated by the eigenfunctions (\ref{dnes})
can be found in~\cite{DIS}.)  Now, the boundary conditions 
(\ref{raogr})--(\ref{bocchiata4}) completely determine the radial 
dependence of the eigensolution of the Schr\"odinger equation
(\ref{rSe0}). Essentially, the regular solution $\chi (r;E)$ is
unique up to multiplication by a function of the energy. Since in Quantum 
Mechanics the boundary conditions (\ref{raogr})--(\ref{bocchiata4}) are 
customarily imposed upon the Schr\"odinger equation, one may be tempted to 
conclude that we have found all the
possible RSPs of the spherical shell potential. As we shall see, this is
not the case: when we multiply $\chi (r;E)$ by the Jost functions, we obtain 
the ``in'' and ``out'' eigensolutions, which generate two new RSPs for the 
spherical shell potential. The ``in'' and ``out'' eigensolutions (and 
therefore their associated RSPs) are determined by the boundary
conditions built into the Lippmann-Schwinger equation.

\subsection{Lippmann-Schwinger Eigenfunctions}
\label{sec:LSe}

The Lippmann-Schwinger equation is usually written as
\begin{equation}
       |E ^{\pm}\rangle =|E\rangle +
       \frac{1}{E-H_0\pm \I \varepsilon}V|E^{\pm}\rangle \, ,
       \label{LSeq1}
\end{equation}
where $H_0$ is the free Hamiltonian. In the radial position representation,
(\ref{LSeq1}) reads
\begin{equation}
       \langle r|E ^{\pm}\rangle =\langle r|E\rangle +
       \langle r|\frac{1}{E-H_0\pm \I \varepsilon}V|E^{\pm}\rangle \, .
       \label{LSeq2}
\end{equation}
The solutions $\langle r|E ^{\pm}\rangle$ of this equation will be called
the ``in'' ($+$) and ``out'' ($-$) Lippmann-Schwinger eigenfunctions. As is 
well known, the integral equation (\ref{LSeq2}) is equivalent to
the Schr\"odinger equation
\begin{equation}
      \left( -\frac{\hbar ^2}{2m} \frac{\D ^2}{\D r^2}+V(r)\right) 
      \langle r|E^{\pm}\rangle =E\langle r|E^{\pm}\rangle 
      \label{selse}
\end{equation}
subject to the following boundary conditions: 
\begin{subeqnarray}
      &&\langle 0|E^{\pm}\rangle =0 \, , \label{bcoLS+-1} \\
      &&\langle r|E^{\pm}\rangle {\rm \ is \ continuous \ at \ } r=a {\rm \
      and \ at \ } r=b \, , \\
      &&\frac{\D }{\D r}\langle r|E^{\pm}\rangle {\rm \ is \ continuous \ at \ } 
       r=a {\rm \ and \ at \ } r=b \, , \label{bcoLS+-3} \\
      && \langle r|E^+\rangle \sim 
          \E ^{-\I kr}- S(k)\E ^{\I kr} \quad {\rm as \ } r\to \infty \, ,
        \label{bcoLS5}\\
      && \langle r|E^-\rangle \sim
         \E ^{\I kr} - S^*(k)\E ^{-\I kr} \quad {\rm as \ } r\to \infty \, ,
        \label{bcoLS6}
\end{subeqnarray}
where $S(k)$ is the $S$ matrix, which is given by the quotient of the Jost 
functions: 
\begin{equation}
       S(E)\equiv S(k)=\frac{{\cal J}_-(k)}{{\cal J}_+(k)} \, .
       \label{smatrix}
\end{equation}
The Jost functions can be written in terms of the coefficients of 
(\ref{chi}) as 
\begin{equation}
      {\cal J}_+(k)=-2i{\cal J}_4(k) \, ; \quad 
      {\cal J}_-(k)=2i{\cal J}_3(k) \, .
        \label{josfuc}
\end{equation}

The $\delta$-normalized [in the sense of (\ref{detanor})] 
Lippmann-Schwinger eigenfunctions can be easily obtained from 
(\ref{selse})--(\ref{josfuc}):
\begin{equation}
      \langle r|E^{\pm}\rangle =\sqrt{\varrho ^{\pm}(k)} \, 
              \frac{\chi (r;k)}{{\cal J}_{\pm}(k)} \, ,
      \label{LSdnes}
\end{equation}
where $\varrho ^{\pm}(k)$ are spectral measures,
\begin{equation}
      \varrho ^+(k)=\varrho ^-(k)=\frac{1}{\pi}\,\frac{2m/\hbar ^2}{k} \, .
      \label{rph}
\end{equation} 
As in the standing-wave case, we are restricting $E$ to $[0,\infty )$. From 
(\ref{smatrix}), (\ref{LSdnes}) and (\ref{rph}) it follows that
the Lippmann-Schwinger eigenfunctions are proportional to each other:
\begin{equation}
      \langle r|E^+\rangle =S(E) \, \langle r|E^-\rangle \, .
       \label{phsfo}
\end{equation}

Applying the Sturm-Liouville theory to $\langle r|E^{\pm}\rangle$ yields 
two other RSPs~\cite{DIS}, i.e., two eigenfunction expansions, two unitary 
operators $U_{\pm}$ that diagonalize the Hamiltonian, two direct integral 
decompositions, and two RHSs\footnote{The RHSs (\ref{rhsofsth}) are 
sketched in~\cite{DIS}. Their complete characterization is a matter of 
current investigation.}
\begin{equation}
      {\mathbf \Phi}_{\pm} \subset {\cal H} \subset 
         {\mathbf \Phi}_{\pm}^{\times} \, . 
      \label{rhsofsth}
\end{equation}
Therefore, even though $\langle r|E\rangle$, $\langle r|E^+\rangle$, 
$\langle r|E^-\rangle$ all fulfill one and the same Schr\"odinger 
equation and have one and the same radial dependence, they generate three 
different
RSPs. Moreover, $\langle r|E^+\rangle$ and $\langle r|E^-\rangle$ differ 
from each other by just a phase factor [see (\ref{phsfo})], because 
$|S(E)|=1$ when $E\in [0,\infty )$. How is then possible that they lead to 
different RSPs? The answer to this question is the following: 
$\langle r|E\rangle$, $\langle r|E^+\rangle$, $\langle r|E^-\rangle$ lead to 
different RSPs because they satisfy different, physically non-equivalent 
boundary conditions.

In order to better understand why $\langle r|E\rangle$, 
$\langle r|E^+\rangle$, $\langle r|E^-\rangle$ yield three different RSPs, we 
compare the Lippmann-Schwinger boundary conditions 
(\ref{bcoLS+-1})--(\ref{bcoLS6}) to the 
standing-wave boundary conditions (\ref{raogr})--(\ref{bocchiata4}). We can 
see first 
that the boundary conditions (\ref{bcoLS+-1})--(\ref{bcoLS+-3}) are the same as
(\ref{raogr})--(\ref{bocchiata4}). We can also see that
in the Lippmann-Schwinger case, we have imposed an additional boundary 
condition that selects the asymptotic behavior of the eigenfunctions
at infinity. For the ``in'' Lippmann-Schwinger eigenfunction, we have chosen
(\ref{bcoLS5}), which means that far away from the potential region 
$\langle r|E^+ \rangle$ is a linear combination of an incoming spherical wave
and an outgoing spherical wave multiplied by the $S$ matrix (which is a
phase factor). For the ``out'' Lippmann-Schwinger eigenfunction, 
we have chosen (\ref{bcoLS6}), which means that far away from the potential 
region $\langle r|E^- \rangle$ is a linear combination of an outgoing 
spherical wave and an incoming spherical wave multiplied by
the complex conjugate of the $S$ matrix (which is also a phase factor). In
the standing-wave case, we did not explicitly impose any boundary 
condition at infinity, which is tantamount to imposing the standing-wave 
asymptotic behavior. Thus $\langle r|E\rangle$, 
$\langle r|E^+\rangle$, $\langle r|E^-\rangle$ differ from each other
just by their asymptotic behavior at infinity. These
different asymptotic behaviors lead, by means of the Jost functions, to 
different analytical properties of the eigensolutions as functions of the 
energy when $E$ is allowed to be complex. Because the Sturm-Liouville 
theory~\cite{DUNFORD} always deals with complex energies, eigenfunctions with 
different analytical properties yield different RSPs. Therefore, what 
ultimately makes $\langle r|E\rangle$, $\langle r|E^+\rangle$, 
$\langle r|E^-\rangle$ different is their different analytical behavior when 
$E$ is allowed to be complex.


An important conclusion can be drawn from the previous paragraph: boundary 
conditions that in the position representation select the asymptotic behavior 
read, in the energy representation, as boundary conditions that select the
analytical behavior, and vice versa. This is particularly apparent in the
Lippmann-Schwinger equation, where the asymptotic boundary conditions
(\ref{bcoLS5}) and (\ref{bcoLS6}) are built into the $\pm \I \varepsilon$ of 
(\ref{LSeq2}), and vice versa. We then say that the 
analytical boundary conditions of the Lippmann-Schwinger equation 
select what is ``in'' ($+\I \varepsilon$) and what is ``out'' 
($-\I \varepsilon$) 
or, equivalently, that the asymptotic behaviors select what is ``in'' 
(\ref{bcoLS5}) and what is ``out'' (\ref{bcoLS6}).

The eigenfunctions $\langle r|E\rangle$, $\langle r|E^+\rangle$, 
$\langle r|E^-\rangle$ all are proportional to the regular solution 
$\chi (r;E)$:
\begin{equation}
      \langle r|E\rangle = \sqrt{\varrho (k)} \, \chi (r;k) \, ,
      \label{dnesb}
\end{equation}
\begin{equation}
      \langle r|E^{+}\rangle =\frac{\sqrt{\varrho ^{+}(k)}}{{\cal J}_{+}(k)} 
                                  \, \chi (r;k) \, ,  \label{LSdnesb+}
\end{equation}
\begin{equation}
      \langle r|E^{-}\rangle =\frac{\sqrt{\varrho ^{-}(k)}}{{\cal J}_{-}(k)} \, 
                               \chi (r;k) \, . \label{LSdnesb-}
\end{equation}
As noted above, the analytical properties of the functions that multiply 
$\chi (r;E)$ in (\ref{dnesb})--(\ref{LSdnesb-}) is what ultimately leads
to different RSPs. However, it is not always true that multiplying 
$\chi (r;E)$ by a function of $E$ yields an eigensolution that generates a 
different RSP. For instance, the radial solution $\chi (r;E)$ and the 
eigenfunction $\langle r|E\rangle$ both lead to the same 
RSP. This is why we say that $\langle r|E\rangle$ is the 
$\delta$-normalization of $\chi (r;E)$. In this case, $\sqrt{\varrho (k)}$ is 
just a normalization factor.

We may then ask: given the regular solution $\chi (r;E)$ and a function of the
energy $f(E)$, how can we know whether $f(E)\chi (r;E)$ is just a
normalization of $\chi (r;E)$ or leads to a different RSP?  A general 
answer to this question is not known. Of course, to know the answer one can 
always apply the Sturm-Liouville theory and see if $f(E)\chi (r;E)$ yields 
the same RSP as $\chi (r;E)$. This may be impractical, though. A faster 
method, that works at least for simple potentials~\cite{DIS,JPA}, is to 
check whether
\begin{equation}
      [f(E^*)]^*=f(E) \, , \quad E\in \mathbb C \, ,
      \label{normcsa1}
\end{equation}
or
\begin{equation}
      [f(E^*)]^*\neq f(E) \, , \quad E\in \mathbb C  \, .
      \label{normcsa2}
\end{equation}     
If $f(E)$ fulfills (\ref{normcsa1}), then $f(E)\chi (r;E)$ is
just a normalization of $\chi (r;E)$. If $f(E)$ fulfills 
(\ref{normcsa2}), then $f(E)\chi (r;E)$ and $\chi (r;E)$ lead to different
RSPs and therefore have different physical content.

To check that the criterion (\ref{normcsa1})--(\ref{normcsa2}) does indeed 
work for the spherical shell potential, we apply it to 
$\langle r|E\rangle$, $\langle r|E^+\rangle$, $\langle r|E^-\rangle$. We 
need first to choose the following branch for the square root function:
\begin{equation}
   \sqrt{\cdot}:\{ E\in {\mathbb C} \, | \  -\pi <{\rm arg}(E)\leq \pi \} 
   \longmapsto \{E\in {\mathbb C} \, | \  -\pi/2 <{\rm arg}(E)\leq \pi/2 \} 
    \, . 
\end{equation}
For $\langle r|E\rangle$, one can easily check that 
\begin{equation}
      [\varrho (E^*)]^*=\varrho (E) \, , \quad E\in \mathbb C \, .
      \label{standw}
\end{equation}
From (\ref{dnesb}) and (\ref{standw}), and from the criterion 
(\ref{normcsa1})--(\ref{normcsa2}) it follows that $\langle r|E\rangle$ is 
just a normalization
of $\chi (r;E)$. For $\langle r|E ^+\rangle$, we have that
\begin{equation}
      [\varrho ^+(E^*)]^*=\varrho ^+(E) \, , \quad E\in \mathbb C \, ,
     \label{rpshi}
\end{equation}      
but
\begin{equation}
      [{\cal J}_+(E^*)]^*={\cal J}_-(E)\neq {\cal J}_+(E) \, , 
      \quad E\in \mathbb C \, . 
       \label{jfuns} 
\end{equation}
From (\ref{LSdnesb+}), (\ref{rpshi}) and (\ref{jfuns}), and from the
criterion (\ref{normcsa1})--(\ref{normcsa2}) it follows that 
$\langle r|E ^+\rangle$ is not a 
normalization of $\chi (r;E)$ but rather has a different physical content. 
Similarly, it can be seen that the physical content of $\langle r|E ^-\rangle$
is not the same as that of $\langle r|E \rangle$.

\subsection{Gamow Eigenfunctions}
\label{sec:GE}

The Gamow vectors are the state vectors of 
resonances~\cite{GAMOW,SIEGERT,MONDRAGON,FERREIRA,AJP,BG,DIS}. Like
the standing-wave and the Lippmann-Schwinger eigensolutions,
they solve the Schr\"odinger equation. At infinity, however, the Gamow 
eigenfunctions satisfy a boundary condition that is different from those
satisfied by $\langle r|E \rangle$, $\langle r|E ^+\rangle$, 
$\langle r|E ^-\rangle$: the purely outgoing boundary condition. This purely 
outgoing behavior determines the physical content of the Gamow vectors.

There are two kinds of Gamow vectors. The first kind is the so-called 
decaying Gamow ket $|z_{\mathrm{R}}^-\rangle$, which is associated to a 
complex energy
$z_{\mathrm{R}}=E_{\mathrm{R}}-\I \Gamma _{\mathrm{R}} /2$ that lies in the 
lower half plane of the second sheet 
of the Riemann surface. The corresponding wave number lies in the fourth 
quadrant of the complex wave-number plane. The second kind of Gamow vector is 
the so-called
growing Gamow ket $|{z_{\mathrm{R}}^*}{^+}\rangle$, which is associated to
an energy $z_{\mathrm{R}}^*=E_{\mathrm{R}}+\I \Gamma _{\mathrm{R}} /2$ that 
lies in the upper half plane of the 
second sheet of the Riemann surface. The corresponding wave number lies in 
the third quadrant of the complex wave-number plane. As far as the 
time-independent Schr\"odinger equation is concerned, any complex number can
be an eigenvalue of the Hamiltonian~\cite{NOTE4}. The role of the purely 
outgoing boundary condition is to select, among all the complex energies, 
those that are to correspond to resonance energies~\cite{AJP}.

In order to obtain the Gamow vectors, we solve the Schr\"odinger differential 
equation
\begin{equation}
       \left( -\frac{\hbar ^2}{2m}\frac{\D ^2}{\D r^2}+V(r)\right) 
       \langle r|z_{\mathrm{R}} \rangle =
       z_{\mathrm{R}} \langle r|z_{\mathrm{R}}\rangle \, ,
	\label{Grse0}
\end{equation}
subject to purely outgoing boundary conditions:
\begin{subeqnarray}
	&&\langle 0|z_{\mathrm{R}}\rangle =0 \label{gvlov1}  \\
	&& \langle r|z_{\mathrm{R}}\rangle \ \mbox{is continuous at} \ r=a \ 
        \mbox{and at} \ r=b    \\
	&&\frac{\D }{\D r}\langle r|z_{\mathrm{R}}\rangle \ \mbox{is continuous at} 
          \ r=a \ \mbox{and at} \ r=b \label{gvlov5} \\
        &&\langle r|z_{\mathrm{R}}\rangle \sim \E ^{\I k_{\mathrm{R}}r} \ \mbox{as} \  r\to \infty \, , 
          \label{gvlov6}  
\end{subeqnarray}
where
\begin{equation}
       k_{\mathrm{R}}=\sqrt{\frac{2m}{\hbar ^2}z_{\mathrm{R}}\,} \, , \quad  
        Q_{\mathrm{R}}=\sqrt{\frac{2m}{\hbar ^2}(z_{\mathrm{R}}-V_0)\,} \, . 
        \label{cmomentum}
\end{equation}
In (\ref{Grse0}) and (\ref{gvlov1})--(\ref{gvlov6}), 
$\langle r|z_{\mathrm{R}}\rangle$ can denote
either $\langle r|z_{\mathrm{R}}^-\rangle$ or $\langle r|{z_{\mathrm{R}}^*}{^+}\rangle$. This will 
cause no confusion, because whenever the complex energy lies in the lower half
plane, $\langle r|z_{\mathrm{R}}\rangle$ will denote $\langle r|z_{\mathrm{R}}^-\rangle$, and
whenever it lies in the upper half plane, $\langle r|z_{\mathrm{R}}\rangle$ will denote 
$\langle r|{z_{\mathrm{R}}^*}{^+}\rangle$.

For the spherical shell potential, (\ref{Grse0}) subject to the 
boundary conditions (\ref{gvlov1})--(\ref{gvlov6}) has solutions only for 
a denumerable set of 
complex energies. These energies come in complex conjugate pairs 
$z_n$, $z_n^*$, where $z_n=E_n -\I \Gamma _n /2$ is the
decaying pole, and $z_n^*=E_n +\I \Gamma _n /2$ is the
growing pole. The corresponding decaying and growing wave numbers
are given by
\begin{equation}
      k_n=\sqrt{\frac{2m}{\hbar ^2}z_n\,} \, , \quad
      -k_n^*=\sqrt{\frac{2m}{\hbar ^2}z_n^*\,} \, , \quad n=1,2, \ldots 
\end{equation}
In terms of the wave number $k_n$, the $n$th decaying Gamow eigensolution 
reads 
\begin{equation}
	\langle r|z_n^-\rangle = N_n\left\{ \begin{array}{ll}
         \frac{1}{{\mathcal J}_3(k_n)}\sin(k_{n}r)  &0<r<a \\ [1ex]
         \frac{{\mathcal J}_1(k_n)}{{\mathcal J}_3(k_n)}\E ^{\I Q_{n}r}
         +\frac{{\mathcal J}_2(k_n)}{{\mathcal J}_3(k_n)}\E ^{-\I Q_{n}r} &a<r<b 
         \\  [1ex]
         \E ^{\I k_{n}r}  &b<r<\infty \, ,
                           \end{array} 
                  \right.
	\label{dgv0p} 
\end{equation}
where $N_n$ is a normalization factor,
\begin{equation}
       N_n^2=\I \, \mbox{res} \left[ S(k) \right]_{k=k_n} \, ,
\end{equation}
and where
\begin{equation}
      Q_n=\sqrt{\frac{2m}{\hbar ^2}(z_n-V_0)\,} \, .
\end{equation}
The $n$th growing Gamow eigensolution reads
\begin{equation}
	\langle r|{z_n^*}{^+} \rangle =M_n\left\{ \begin{array}{ll}
         \frac{1}{{\mathcal J}_3(-k_n^*)}\sin(-k_{n}^*r)  &0<r<a \\ [1ex]
         \frac{{\mathcal J}_1(-k_n^*)}{{\mathcal J}_3(-k_n^*)}\E ^{\I Q_{n}^*r}
         +\frac{{\mathcal J}_2(-k_n^*)}{{\mathcal J}_3(-k_n^*)}
         \E ^{-\I Q_{n}^*r} &a<r<b \\ [1ex]
         \E ^{-\I k_{n}^*r}  &b<r<\infty \, .
                           \end{array} 
                  \right.
	\label{ggv0p} 
\end{equation}
where $M_n$ is a normalization factor,
\begin{equation}
       M_n^2=\I \, \mbox{res} \left[ S(k) \right]_{k=-k_n^*}=(N_n^2)^*  \, ,
\end{equation}
and where
\begin{equation}
       Q_n^*=\sqrt{\frac{2m}{\hbar ^2}(z_n^*-V_0)\,} \, .
\end{equation}

We compare now the boundary conditions (\ref{gvlov1})--(\ref{gvlov6}) 
satisfied by the Gamow 
eigenfunctions with those satisfied by the standing-wave eigenfunctions 
[see (\ref{raogr})--(\ref{bocchiata4})] and by the Lippmann-Schwinger 
eigenfunctions 
[see (\ref{bcoLS+-1})--(\ref{bcoLS6})]. Clearly, the boundary condition 
that is specific 
to the Gamow vectors is (\ref{gvlov6}). This boundary condition singles out 
the complex resonance spectrum of the Schr\"odinger equation by specifying 
the asymptotic behavior of the Gamow eigenfunctions. Moreover, the
resonance spectrum selected by (\ref{gvlov6}) coincides with the poles of the 
$S$ matrix (\ref{smatrix}). 

To finish this section, we note that in the $S$-matrix formalism, a resonance
energy is reached by analytic continuation of $S(E)$ from its values on the 
physical spectrum to the resonance pole. In a similar vein, Gamow vectors can
be viewed as the solutions of the analytic continuation of the Schr\"odinger 
equation [subject to the boundary conditions (\ref{gvlov1})--(\ref{gvlov6})] 
from the energies
in the physical spectrum to the complex resonance eigenvalue.

\section{The Arrow of Time of Quantum Electrodynamics}
\label{sec:tdse}

We have seen how boundary conditions determine the physical content of the
eigensolutions of the time-independent Schr\"odinger equation. We now turn
to see how boundary conditions affect the physical content of the solutions
of the time-dependent Schr\"odinger equation. We shall see that 
time-asymmetric boundary conditions imposed upon the (time-symmetric) 
Schr\"odinger equation yield solutions that have an arrow of time built into 
them. Quantum Electrodynamics (QED) will be used as an illustrative example.

In order to make a parallel with the quantum case, we recall first the 
essential features of the arrow of time of Classical Electrodynamics 
(CED). The Maxwell equations, which
describe the classical electromagnetic fields, are 
time symmetric. The solutions of the Maxwell equations can be written
as a combination of a retarded and an advanced 
solution. Experimentally, we always observe that light has a retarded
behavior -- light cannot be detected at a distance $R$ from the source at
any time less than $R/c$. In order to account for this retarded behavior, we
select the retarded solution of the Maxwell equation and forbid the
advanced solution. This amounts to imposing a time-asymmetric, retarded, 
causal boundary condition upon the (time-symmetric) Maxwell equations. 
Essentially, this is the radiation arrow of time. We stress that this arrow of 
time stems ultimately from the imposition of a causal boundary condition.

QED is the quantum counterpart of CED, and also has an arrow of time built 
into it. The arrow of time of QED is built into the
Feynman propagator. Although the Schr\"odinger equation is 
time symmetric, one always imposes Feynman's $+\I \varepsilon$ prescription to 
construct the Feynman propagator -- particles travel forward in time, whereas 
antiparticles ``travel backward'' in time~\cite{FEYNMAN}. Thus Feynman's 
$+\I \varepsilon$ prescription imposes a retarded (advanced) condition on the 
time evolution of particles (antiparticles). Essentially, this 
is the arrow of time of QED. In particular, this arrow of time sheds light 
onto the physical meaning of Feynman's $+\I \varepsilon$ prescription: this 
prescription is just a causal boundary condition imposed upon the time 
evolution of particles and antiparticles. 

The example of QED can be generalized to any closed quantum system. Although
the Schr\"odinger equation (which describes the evolution of a closed 
quantum system) is time symmetric, physical processes seem to
comply with our basic notions of causality: cause is prior to effect. In 
order to 
make the solutions of the (time-symmetric) Schr\"odinger equation comply with 
causality, we impose causal, time-asymmetric boundary conditions upon the
Schr\"odinger equation. These boundary conditions single out the causal 
solutions that account for the observed time-asymmetric phenomena. Building 
that (e.g., retarded) boundary condition into the time evolution of the 
quantum system involves the construction of an (e.g., retarded) 
propagator. Actually, although it is well known that propagators have an 
arrow of time built into them, it is not so well emphasized~\cite{HALLIWEL} 
that this implies the existence of a fundamental time asymmetry at the 
microscopic level.

Many authors have realized the central
role played by boundary conditions in the description of time
asymmetry. For example, Ritz~\cite{RITZ}
thought that the time asymmetry (often called irreversibility) of statistical 
mechanics arises from boundary conditions (in contrast to 
Einstein~\cite{RITZ}, who thought that irreversibility comes from 
averaging over a large number of systems, that is, for Einstein irreversibility
emerges from probability and statistics). For Peierls~\cite{PEIERLS} 
irreversibility also arises as a consequence of boundary conditions. To show
this, Peierls rephrases Boltzmann's Stosszahl-Ansatz, which is 
the origin of the irreversibility of the so-called ``Lorentz gas,'' as 
a boundary condition~\cite{PEIERLS}. Some authors such as 
Penrose~\cite{PENROSE} or Gell-Mann and Hartle~\cite{GH} have used boundary 
conditions as a possible explanation of time asymmetry, and even of time 
symmetry~\cite{GH}. Other authors such as Preskill~\cite{PRESKILL} also
use boundary conditions to explain the time asymmetry (irreversibility) of 
quantum statistical mechanics: essentially, the non-decreasing 
behavior of the entropy can be understood as stemming from
the assumption that system and environment are initially uncorrelated, i.e.,
from the assumption that the initial system-environment state is separable 
(unentangled)~\cite{PRESKILL}.

\section{Conclusions}
\label{sec:conclusions}

We have seen how important boundary conditions are in obtaining the solutions 
of the time-independent Schr\"odinger equation that fit a given physical 
situation. Essentially, the asymptotic behavior of the eigensolution 
determines its physical content. We have analyzed and compared the 
standing-wave, ``in,'' ``out,'' and purely outgoing boundary 
conditions. We have seen that the standing-wave, ``in,'' and ``out'' 
boundary conditions yield three physically different RSPs. The 
purely outgoing boundary condition selects the resonance spectrum.

We have discussed the time-asymmetry of QED. Essentially, Feynman's
$+\I \varepsilon$ prescription, which is used to construct the Feynman 
propagator,
encodes the time asymmetry of QED. We have also 
concluded that in general, the time evolution of a closed quantum system
has a time asymmetry built into the propagators.

\section*{Acknowledgment}

The author thanks F.~Gaioli for drawing his attention to the time asymmetry of
the Feynman propagator, and W.~H.~Zurek for a stimulating discussion on
the arrow of time. The discussions with F.~Gaioli and W.~H.~Zurek led in part 
to Section~\ref{sec:tdse}. The author received, once again, invaluable 
English-style advise from C.~Koeninger, to whom the author is very 
grateful. 

This work was financially supported by the E.U.~TMR Contract 
No.~ERBFMRX-CT96-0087 ``The Physics of Quantum Information.''

%


\begin{thebibliography}{8.}
\bibitem{RSP} Obviously, the Hamiltonian has more RHS properties than those
included in our definition of RSP.
\bibitem{NOTE4} This is in no contradiction with the self-adjointness of the
Hamiltonian, because the eigensolutions that correspond to complex eigenvalues
lie outside the Hilbert space.
\end{thebibliography}
\end{document}